\documentclass[epj]{svjour}
\usepackage{amsmath}
\usepackage{amssymb}
\usepackage{latexsym}
\usepackage{graphics}
\usepackage{graphicx}
\usepackage{epsfig}
\usepackage{epsf}

\def\slash#1{\setbox0=\hbox{$#1$}
   \dimen0=\wd0 \setbox1=\hbox{/} \dimen1=\wd1
   \ifdim\dimen0>\dimen1 \rlap{\hbox to \dimen0{\hfil/\hfil}} #1
   \else  \rlap{\hbox to \dimen1{\hfil$#1$\hfil}} / \fi}
\def\intk{\int \frac{d^4 k}{(2\pi)^4}}

\begin{document}

\title{Deuteron radial moments for renormalized chiral
potentials} \author{\underline{E. Ruiz Arriola} 
%\inst{1}
\thanks{\emph{Poster presented at QNP06 Madrid, 5-10 June 2006. Work
supported by Spanish DGI and FEDER funds with grant no. BFM2002-03218,
Junta de Andaluc\'{\i}a grant No. FQM-225, and EU RTN Contract
CT2002-0311 (EURIDICE) }} and M. Pavon Valderrama 
%\inst{1}
% etc
}                     % Do not remove
\institute{
Departamento de F\'{\i}sica At\'omica, Molecular y
Nuclear, Universidad de Granada, E-18071 Granada, Spain
}
\date{\today}
%\date{Oct. 1, 2006}
\abstract{We calculate deuteron positive and negative radial moments 
involving any bilinear function of the deuteron S and D wave functions 
for renormalized OPE and TPE chiral potentials. The role played by the
strong singularities of the potentials at the origin and the short 
distance insensitivity of the results when the potentials are fully 
iterated is emphasized  as compared to realistic potentials. 
\PACS{
      {03.65.Nk}{11.10}  
      {13.75}
     } % end of PACS codes
} %end of abstract
\maketitle

Chiral dynamics has played an important role in the theoretical
description of low energy hadronic reactions~\cite{Ericson:1988gk} and
so far is the only known vestige of the underlying fundamental QCD
theory of strong interactions in nuclear physics.  There is a number
of low energy theorems based on chiral symmetry which provide a
quantitative and model independent insight into low energy processes
involving pions and nucleons, due to the clear scale separation
between nuclear physics and QCD.  For compound systems which at low
energies disclose their composite nature the theoretical description
necessarily becomes very involved and probably dependent on arbitrary
assumptions. On the contrary, for weakly bound systems such as the
deuterium nucleus one expects important simplifications leading to a
more scheme independent and possibly systematic description of these
systems. This possibility motivated the introduction of Effective
Field Theory (EFT) approaches~\cite{Weinberg:1990rz} for nuclear
physics based on the chiral symmetry of QCD, and the derivation of low
energy theorems, as, for example, pion-deuteron
scattering~\cite{Weinberg:1992yk} (for comprehensive reviews see
e.g. Ref.~\cite{Bedaque:2002mn,Phillips:2002da,Phillips:2005vv}).  In
many cases most of the information needed for reactions involving the
deuteron can be encoded by simple deuteron matrix elements.

Guided by earlier work~\cite{Martorell94,Beane:2001bc}, we have
proposed~\cite{PavonValderrama:2003np,PavonValderrama:2004nb,PavonValderrama:2005gu,Valderrama:2005wv,PavonValderrama:2005uj}
to renormalize the NN interaction in a non-perturbative way,
highlighting model independent long distance correlations among
physical observables.  In our approach the long distance chiral NN One
Pion Exchange (OPE) and Two Pion Exchange (TPE) potentials, computed
within perturbation theory in
Refs.~\cite{Kaiser:1997mw,Friar:1999sj,Rentmeester:1999vw}, are
iterated to all orders in the Schr\"odinger equation very much in the
spirit of the original Weinberg
approach~\cite{Weinberg:1990rz}. However, some subtleties
are
found~\cite{PavonValderrama:2004nb,Valderrama:2005wv,Nogga:2005hy},
which impose strong constraints on the admisible short distance
physics based on orthogonality, uniqueness and finiteness of the results.
\begin{figure*}[ttt]
\begin{center}
\epsfig{figure=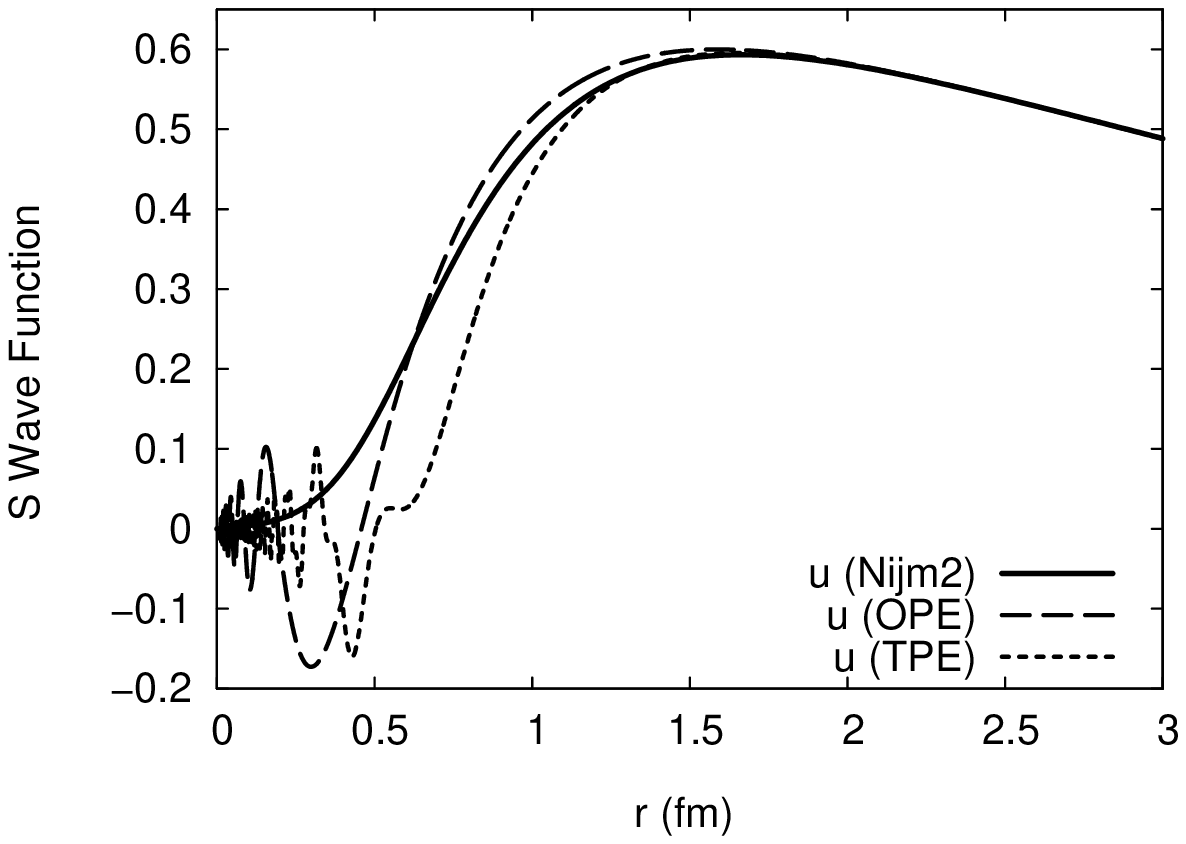,height=6cm,width=8cm}
\epsfig{figure=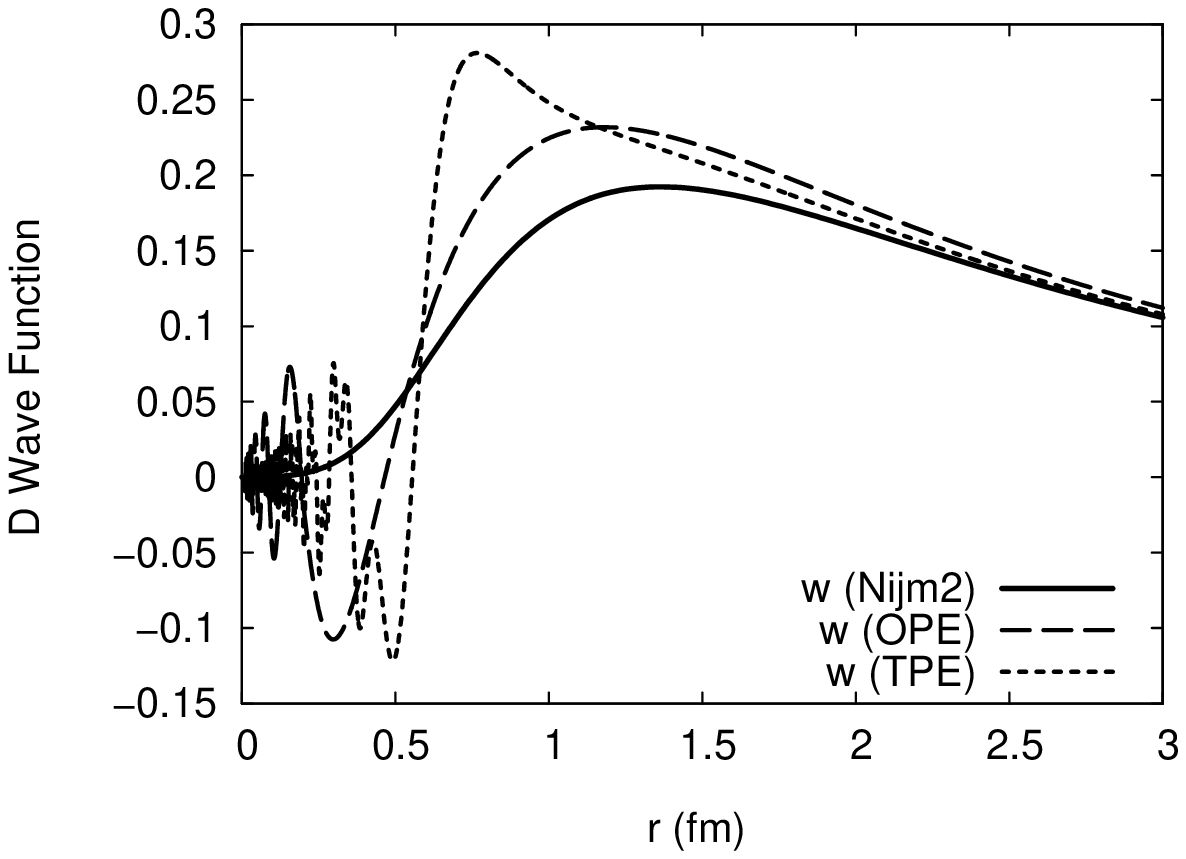,height=6cm,width=8cm}
\end{center}
\caption{The OPE and TPE deuteron wave functions, u (left) and
w (right), as a function of the distance (in {\rm fm}) compared to the
Nijmegen II wave functions~\cite{Stoks:1994wp}. The asymptotic normalization 
$u \to e^{-\gamma r}$ has been adopted and the asymptotic D/S ratio is 
taken $\eta = 0.0256 (4)$ in the TPE case (for OPE $\eta = 0.026333$). 
We use the set IV of chiral couplings (see Ref.~\cite{Valderrama:2005wv}).}
\label{fig:u+w_TPE}
\end{figure*} 

In the $^3S_1-{}^3D_1$ channel, the relative proton-neutron state 
for negative energy is described by the coupled equations
\begin{eqnarray}
\begin{pmatrix}
- \frac{d^2}{dr^2} + U_s (r) & U_{sd} (r) \\
U_{sd}(r) & - \frac{d^2}{dr^2} + \frac{6}{r^2} + U_d (r) 
\end{pmatrix}
\begin{pmatrix}
u \\ w  
\end{pmatrix} 
= - \gamma^2 
\begin{pmatrix}
u \\ w  
\end{pmatrix} 
\label{eq:sch_coupled} 
\end{eqnarray}
Here $ \gamma = \sqrt{M B} $, with $B$ the deuteron binding energy
and $M$ the nucleon mass, $ U (r) = M V(r) $ are the reduced
potentials and $u (r)$ and $w (r)$ are S- and D-wave deuteron reduced
wave functions respectively. At long distances they satisfy,
\begin{eqnarray}
\begin{pmatrix} 
u \\ 
w 
\end{pmatrix} \to  A_S\,e^{-\gamma r} \, 
\begin{pmatrix}
1 \\ 
\eta \left[1 + \frac{3}{\gamma r} + \frac{3}{(\gamma r)^2} \right] 
\end{pmatrix} 
\label{eq:bcinfty_coupled} 
\end{eqnarray}
where $\eta$ is the asymptotic D/S ratio parameter and $A_S$ is the
asymptotic normalization factor, which is such that the deuteron wave 
functions are normalized to unity.
For conventions and numerical values of parameters we use
Ref.~\cite{PavonValderrama:2005gu,Valderrama:2005wv} throughout.  

In this work we report on the radial moments
\begin{eqnarray}
\langle r^n \rangle_u &=& \int_0^\infty  r^n u(r)^2 dr \\   
\langle r^n \rangle_w &=& \int_0^\infty  r^n w(r)^2 dr  \\ 
\langle r^n \rangle_{uw} &=& \int_0^\infty  r^n u(r) w(r) dr  
\end{eqnarray} 
for $-3 \le n \le 2 $ which appear in many situations of interest, 
such as the calculation of the matter radius, the deuteron quadrupole moment 
and deuteron magnetic moment for the positive powers, as well as $\pi d $ 
and $K d $ elastic scattering and neutral pion photoproduction, $\gamma d \to
\pi^0 d$, in the case of the negative powers. 

An important issue is the finiteness of the negative radial moments, a
topic which has been recently discussed for
OPE~\cite{PavonValderrama:2005gu,Nogga:2005fv,Platter:2006pt} and
TPE~\cite{Valderrama:2006np}. The remarkable finding is that chiral
potentials~\cite{Kaiser:1997mw,Friar:1999sj,Rentmeester:1999vw}, when
fully iterated, have an increasing number of finite inverse radial
moments due to the near the origin singularities of the
potential. They smoothen the short distance behaviour of the wave functions
and hence improve the convergence of the inverse radial moments.  This
is in sharp contrast with perturbative
approaches~\cite{Borasoy:2003gf}, for which the perturbative wave
functions diverge at the
origin~\cite{PavonValderrama:2005gu,Valderrama:2005wv}~\footnote{This
divergency holds for perturbations both on boundary conditions or on
distorded (fully iterated) OPE waves.}, or conventional (regular)
phenomenological potentials~\cite{Stoks:1994wp} where the S- and
D-wave short distance behaviour of the wave functions, $u \sim r $ and $w
\sim r^5 $ respectively, is enough to render $\langle 1/r \rangle_u$
and $\langle 1/r^2 \rangle_u$ finite, but produce divergent higher
inverse moments.  We illustrate the situation below.

At distances much shorter than the pion Compton wavelength, 
the OPE potential behaves as
\begin{eqnarray} 
\begin{pmatrix} 
U_{s}^{\rm OPE} (r) &  U_{sd}^{\rm OPE} (r) \\ 
U_{sd}^{\rm OPE} (r) &  U_{d}^{\rm OPE} (r) 
\end{pmatrix} 
\to \frac{1}{r^3} 
\begin{pmatrix} 
R_{s}  & R_{sd} \\  
R_{sd}  & R_{d}
\end{pmatrix} 
\end{eqnarray} 
with $R_s = 0$, $R_{sd}= 2 \sqrt{2} R $, $R_d = 4 R$ and  
$R = 3 g_A^2 M / 32 \pi f^2$ ($= 1.07764\,{\rm fm}$). 
This behaviour of the potential is strong enough to overcome the centrifugal
barrier at short distances, thus modifying the usual short distance behaviour of 
the wave functions, which can schematically be written as
\begin{eqnarray}
u(r) \sim w(r) \sim \left(\frac{r}{R}\right)^{3/4} f\left(\frac{r}{R}\right)
\end{eqnarray}
where $f(r/R)$ represents some linear combination of \break $\sin{(4
\sqrt{R/r})}$, $\cos{(4 \sqrt{R/r})}$ and $\exp{(-4
\sqrt{2}\sqrt{R/r})}$ (for a complete analysis see
Ref.~\cite{PavonValderrama:2005gu}). The elimination of the diverging
exponential fixes $\eta_{\rm OPE}=0.0263$.  From this short distance
behaviour of the wave functions, one finds that the $\langle 1/r
\rangle_u$ and $\langle 1/r^2 \rangle_u$ moments are finite for the
OPE potential, while $\langle 1/r^3 \rangle_u$ and higher moments
diverge, as it would happen for a regular potential.

The short distance behaviour of the TPE (NNLO) has been exploited in
Ref.~\cite{Valderrama:2005wv}.  The potential at short distances
behaves as~\cite{Kaiser:1997mw,Friar:1999sj,Rentmeester:1999vw}
\begin{eqnarray}\label{eq:U_TPE_matrix} 
\begin{pmatrix} 
U_{s}^{\rm TPE} (r) &  U_{sd}^{\rm TPE} (r) \\ 
U_{sd}^{\rm TPE} (r) &  U_{d}^{\rm TPE} (r) 
\end{pmatrix} 
\to \frac{1}{r^6} 
\begin{pmatrix} 
R_{s}^4  & R_{sd}^4 \\  
R_{sd}^4  & R_{d}^4
\end{pmatrix} 
\end{eqnarray} 
where 
\begin{eqnarray} 
(R_{s})^4 &=& \frac{3 g_A^2}{128 f^4 \pi^2 } ( 4 - 3 g_A^2 + 24 \bar
c_3 - 8 \bar c_4 ) \nonumber \\ (R_{sd})^4 &=& - \frac{3 \sqrt{2}
g_A^2}{128 f^4 \pi^2 } (-4 + 3 g_A^2 - 16 \bar c_4 ) \nonumber \\
(R_{d})^4 &=& \frac{9 g_A^2}{32 f^4 \pi^2 } (-1+2 g_A^2 + 2 \bar c_3 -
2 \bar c_4 ) 
\label{eq:vdw_triplet}
\end{eqnarray} 
and $ \bar c_i = M c_i$ are the low energy chiral couplings appearing
in $\pi N $ scattering.
As in the OPE case, this potential is strong enough at short distances 
to modify the short distance behaviour of the wave function, which now reads
\begin{eqnarray}
u(r) \sim w(r) &\sim& 
C_{+}\,\left({r}/{R_+}\right)^{3/2} f_+\left({r}/{R_+}\right) + 
\nonumber\\
&& C_{-}\,\left({r}/{R_-}\right)^{3/2} f_-\left({r}/{R_-}\right)
\end{eqnarray}
where $R_+^4$ and $R_-^4$ are the eigenvalues of the matrix in
Eq.~(\ref{eq:U_TPE_matrix}), and $f_{\pm}(r/R_{\pm})$ represents a
linear combination of $\sin{({R_{\pm}^2/2 r^2})}$ and
$\cos{({R_{\pm}^2/2 r^2})}$ leaving $\eta_{\rm TPE}$ as a free
parameter~\cite{Valderrama:2005wv,Valderrama:2006np}.  From this short
distance behaviour, the $\langle 1/r \rangle_u$, $\langle 1/r^2 \rangle_u$
and $\langle 1/r^3 \rangle_u$ radial moments are finite for the TPE
potential, while higher moments diverge (although they would become
finite for higher order potentials).

The wave functions for OPE and TPE as compared to the NijmII ones have
been depicted in Fig.~\ref{fig:u+w_TPE}. The radial moments are
tabulated in Table~\ref{tab:r_moments}. As we see, and despite the
very different behaviour at short distances between the deuteron wave
functions corresponding to renormalized chiral potentials and to
phenomenological potentials, the convergent moments are fairly similar
(the u+w combination works better) despite that NijmII and Reid93
contain no explicit TPE components. For inverse moments the trend
improves clearly when going from OPE to TPE, which we interpret as a
correct implementation of model independent long distance correlations
generated by chiral symmetry and renormalization constraints.
\begin{table*}[ttt]
\begin{center}
\begin{tabular}{|c|c|c|c|c|c|c|c|c|c|c|}
\hline \hline & Short & OPE & OPE${}^*$ & TPE-SetI & TPE-SetII &
                TPE-SetIII & TPE-SetIV & NijmII & Reid93 \\ \hline
$\gamma\,({\rm fm}^{-1})$ & Input &  Input & Input & Input & Input & Input 
& Input & 0.231605 & 0.231605 \\
$\eta$ & 0.0 & 0.026333 & 0.025547 & Input & Input & Input 
& Input & 0.02521 & 0.02514  \\ 
\hline 
$\langle r^{2} \rangle_u  $     & 9.3213 & 14.582(6) & 14.424(6) & 15.60(9) 
& 15.61(11) & 15.3(3) & 15.09(13) & 15.129 & 15.147 \\ 
$\langle r^{2} \rangle_w  $     & 0.0 & 0.3849(2) & 0.36371(15) & 0.37(3) 
& 0.38(2) & 0.37(2) & 0.38(2) & 0.3438 & 0.3429 \\ 
$\langle r^{2} \rangle_{uw}  $  & 0.0 & 2.0883(9) & 2.0144(8) & 2.14(4) 
& 2.15(3) & 2.11(3) & 2.09(3) & 2.035  & 2.032 \\ 
\hline 
$\langle r^{1} \rangle_u  $    & 2.1589 & 3.0400(14) & 3.0185(13) & 3.21(2) 
& 3.20(2) & 3.16(2) & 3.12(3) & 3.138 &  3.139 \\ 
$\langle r^{1} \rangle_w  $    & 0.0 & 0.14287(6) & 0.13691(6) & 0.134(12) 
& 0.139(12) & 0.136(12) & 0.146(12) & 0.1204 & 0.1206 \\ 
$\langle r^{1} \rangle_{uw}  $ & 0.0 & 0.5898(3)  & 0.5739(2) & 0.586(14) 
& 0.589(14) & 0.581(15) & 0.584(13) & 0.5594 & 0.5590 \\ 
\hline 
$\langle r^{0} \rangle_u  $    & 1.0 & 0.9270(4)  & 0.9287(4) & 0.935(9) 
& 0.930(9) & 0.931(10) & 0.918(10) & 0.9436  & 0.9430 \\ 
$\langle r^{0} \rangle_w  $    & 0.0 & 0.07312(3) & 0.07146(2) & 0.065(9) 
& 0.070(9) & 0.069(10) & 0.081(10) & 0.05635 & 0.05699 \\ 
$\langle r^{0} \rangle_{uw}  $ & 0.0 & 0.23989(11)& 0.23691(10) & 0.222(7) 
& 0.225(7) & 0.225(8) & 0.233(7) & 0.2166  & 0.2172 \\ 
\hline 
$\langle r^{-1} \rangle_u $    & $\infty$ & 0.4259(3)  & 0.4336(2) & 0.382(5) 
& 0.377(5) & 0.388(6) & 0.384(5) & 0.4160  & 0.4163 \\  
$\langle r^{-1} \rangle_w $    & 0.0 & 0.052498(5)& 0.05256(3) & 0.042(8) 
& 0.048(7) & 0.048(10) & 0.063(10) & 0.03419 & 0.03520 \\ 
$\langle r^{-1} \rangle_{uw} $ & 0.0 & 0.14120(7) & 0.14239(3) & 0.112(5) 
& 0.115(5) & 0.117(6) & 0.128(6) & 0.1153  & 0.1166 \\  
\hline 
$\langle r^{-2} \rangle_u $    & $\infty$ & 0.3464(8) & 0.3582(3) & 0.210(4) 
& 0.205(4) & 0.220(5) & 0.221(3) & 0.2607  & 0.2646 \\ 
$\langle r^{-2} \rangle_w $    & 0.0 & 0.0771(2) & 0.0783(3) & 0.038(8) 
& 0.044(7) & 0.045(12) & 0.064(11) & 0.02613 & 0.02780 \\ 
$\langle r^{-2} \rangle_{uw} $ & 0.0 & 0.1551(4) & 0.1589(3) & 0.072(4) 
& 0.075(4) & 0.079(4) & 0.093(6) & 0.08122  & 0.08413 \\ 
\hline 
$\langle r^{-3} \rangle_u $    & $\infty$ & $\infty$ & $\infty$ & 0.159(3)
& 0.155(3) & 0.173(4) & 0.1851(8) & $\infty$ & $\infty$ \\
$\langle r^{-3} \rangle_w $    & 0.0 & $\infty$ & $\infty$ & 0.053(10)
& 0.059(9) & 0.066(14) & 0.091(14) & 0.02465 & 0.02783 \\
$\langle r^{-3} \rangle_{uw} $ & 0.0 & $\infty$ & $\infty$ & 0.0626(14)
& 0.068(2) & 0.071(2) & 0.097(6) & 0.07342 & 0.08064 \\
\hline 
\hline
\end{tabular}
\end{center}
\caption{Deuteron radial moments (in units of powers of fm).
We consider the OPE and TPE potentials; in the case of the OPE potential 
we have taken $g_{\pi NN} = 13.08$ (i.e. $g_A = 1.29$, OPE) and 
$g_A = 1.26$ (OPE$^*$), while in the TPE case we show the results 
corresponding to the four set of chiral couplings considered along our previous
works~\cite{Valderrama:2005wv,Valderrama:2006np}.
In the OPE case the error is estimated by varying the semiclassical matching 
radius~\cite{PavonValderrama:2005gu,Valderrama:2006np} 
in the $0.1 - 0.2\,{\rm fm}$ range , while in the TPE case the error 
comes from the experimental uncertainty of the $D/S$ ratio, $\eta = 0.0256(4)$.
TPE Sets I,II,II and IV refer to the chiral parameters, $c_1$, $c_3$ and
$c_4$ of Refs.~\cite{Buettiker:1999ap}, \cite{Rentmeester:1999vw},
\cite{Entem:2002sf} and \cite{Entem:2003ft} respectively. 
Nijm II and Reid93 are calculated from Ref.~\cite{Stoks:1994wp} or taken from 
Ref.~\cite{deSwart:1995ui}.}
\label{tab:r_moments}
\end{table*}

\end{document}